\documentclass[cup6a]{cupbook}
\usepackage{epsfig}
\title[ AGN Feedback in Galaxy Formation]
      {Proceedings of the Workshop held in Vulcano (Messina), Italy, May 18-22, 2008}
\author{V.Antonuccio-Delogu and J. Silk, eds. \TeX-to-type}
\date{\today}

\begin{document}


\chapter[Giant outflows in Powerful z$\sim$ 2 Radio Galaxies]{Giant Outflows in powerful z$\sim$2 Galaxies: The “Smoking Gun” of AGN Feedback in the Early Universe}

\section{Introduction}
AGN feedback is now a critical component of models of galaxy
formation, introduced to bring several outstanding issues of our
hierarchical picture of galaxy evolution into agreement with
observations \cite{benson03,croton06}. The energy ejected by powerful
AGN is similar to the binding energy of the massive host galaxy and
may therefore {\it in principle} be sufficient to offset cooling and
star formation in galaxies \cite{silk98,antonuccio08}. However, it is not clear by
what mechanism this energy is being transformed into kinetic energy of
the gas.

Observationally a picture emerges where this energy transfer may occurc
mostly through the relativistic, synchrotron-emitting plasma ejected
by radio-loud AGN
\cite{mcnamara07,best06,holt08}. However, most of these observations
focus on galaxies at low redshifts, whereas the main formation phase
of massive galaxies was in the early Universe, at redshifts z
$\sim$2. Largely driven by the evolution of the host galaxy and its
surroundings, the role of AGN in regulating star formation may be
different in galaxies at high redshift compared to galaxies
today. At high redshift, the energy injected into the interstellar
medium may have contributed to ``quenching'' the vigorous starbursts
during the main formation phase of a massive galaxy, by heating and
removing the molecular reservoirs of star formation (``quenching
phase''). At low redshift, subsequent, perhaps repeated and weaker
phases of AGN activity may play a role in inhibiting further gas
cooling over significant cosmic times (``maintenance phase''). 

\section{Giant outflows in powerful radio galaxies at z$\sim$2}
Powerful radio galaxies are ideal targets to search for the
fingerprints of AGN feedback in the early Universe. They are
particularly massive \cite{debreuck02,seymour07} and as a population
are likely undergoing a transition from strongly star forming to 'old,
red, and dead' \cite{reuland04,archibald01}. They are
also the hosts of particularly powerful, radio-loud AGN. Deep Ly$\alpha$
spectroscopy, including integral-field spectroscopy, shows that many radio
galaxies are embedded into Mpc-sized gaseous reservoirs, perhaps the
'vestiges' from which the radio galaxy initially formed
\cite{villar02,villar03,villar06,villar07,humphrey07}. 

Using rest-frame optical integral-field spectroscopy at the VLT, we identified
spatially extended, kpc-sized outflows of ionized gas which correspond to
significant fractions of the interstellar medium of a massive, gas-rich galaxy
\cite{nesvadba06,nesvadba07,nesvadba08}, and which have about solar
metallicity \cite{humphrey08b}. Based on the observed H$\alpha$ emission-line
luminosities, electron densities measured from the
[SII]$\lambda\lambda$6716,6731 line ratio, and simple case~B recombination, we
estimate gas masses of ionized gas of everal $10^{10}$ M$_{\odot}$. This
corresponds to the molecular gas masses in the strongest starburst galaxies at
similar redshifts. This is also similar, or even exceeds, the molecular gas
masses in those galaxies of our sample for which we have direct estimates of
the cold molecular gas mass measured from the CO line emission. This may
suggest that the outflows indeed include a significant fraction of the
interstellar medium of these galaxies.

Using imaging spectroscopy, we also infer the kinematics and kinetic energy of
the gas. We find relatively uniform velocity patterns with one strongly
redshifted and one strongly blueshifted bubble suggestive of a biploar flow
(Fig.~1). Measured velocity offsets are of order 1000 km s$^{-1}$ between
bubbles, not correcting for inclination. Line widths of FWHM$\sim$800-1000 km
s$^{-1}$ suggest this gas is very turbulent.

\begin{figure}
\epsfig{figure=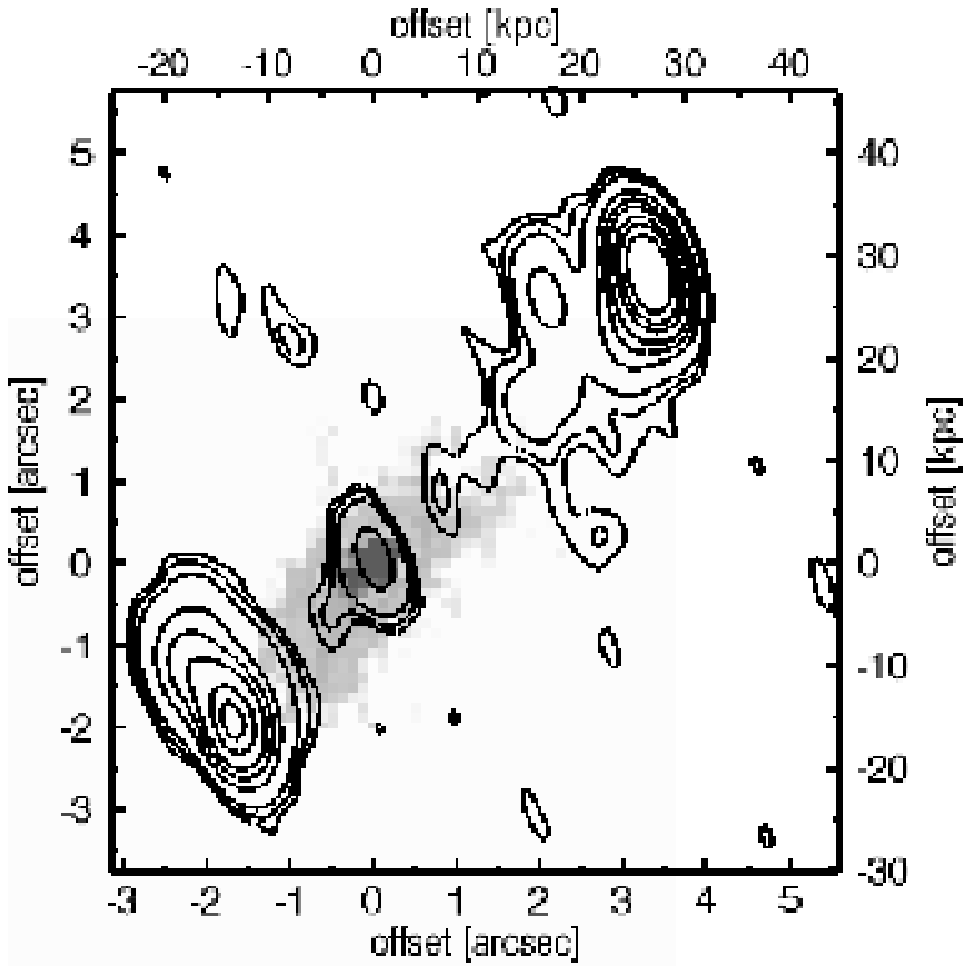,width=0.28\textwidth}
\epsfig{figure=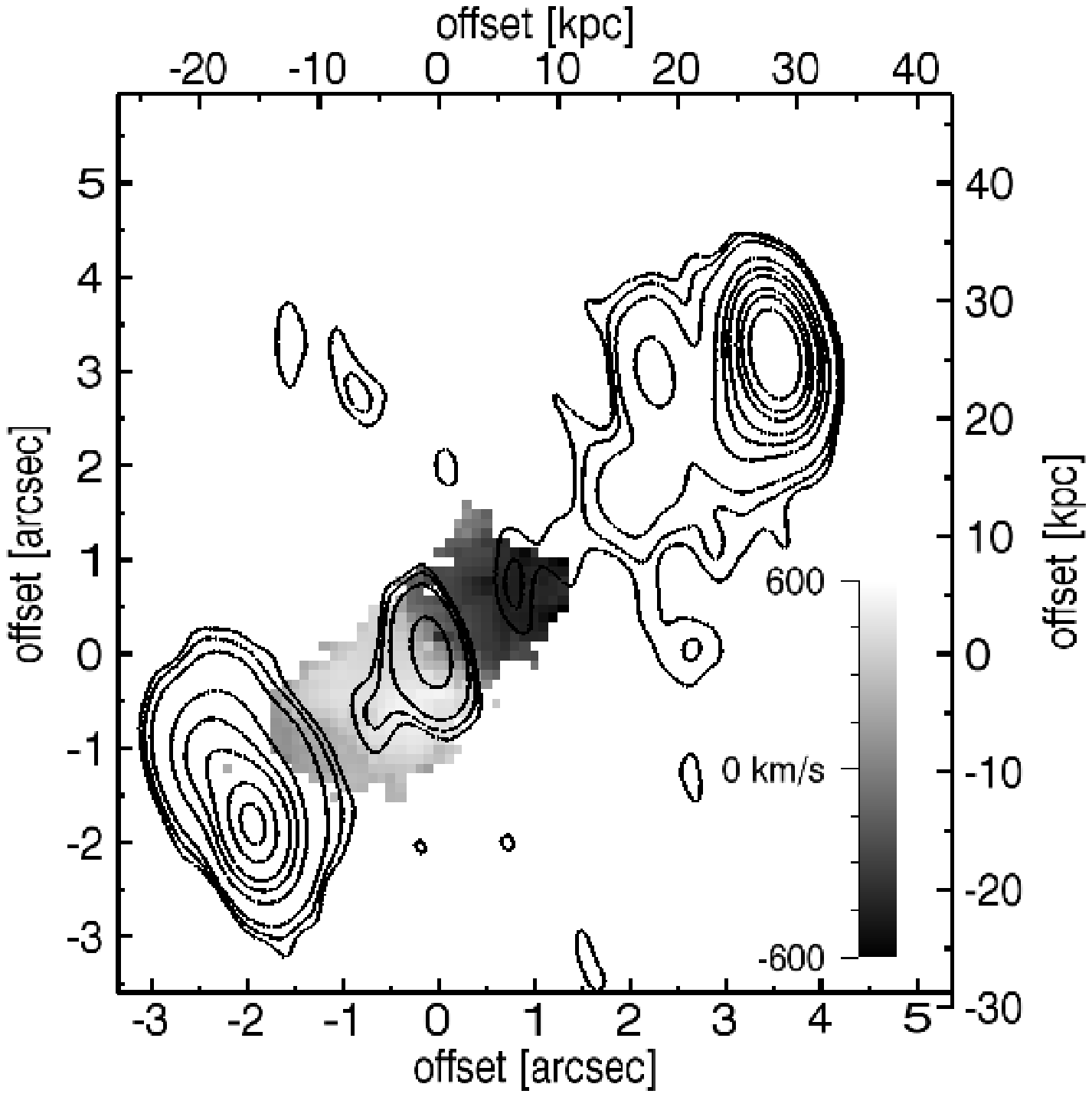,width=0.28\textwidth}
\epsfig{figure=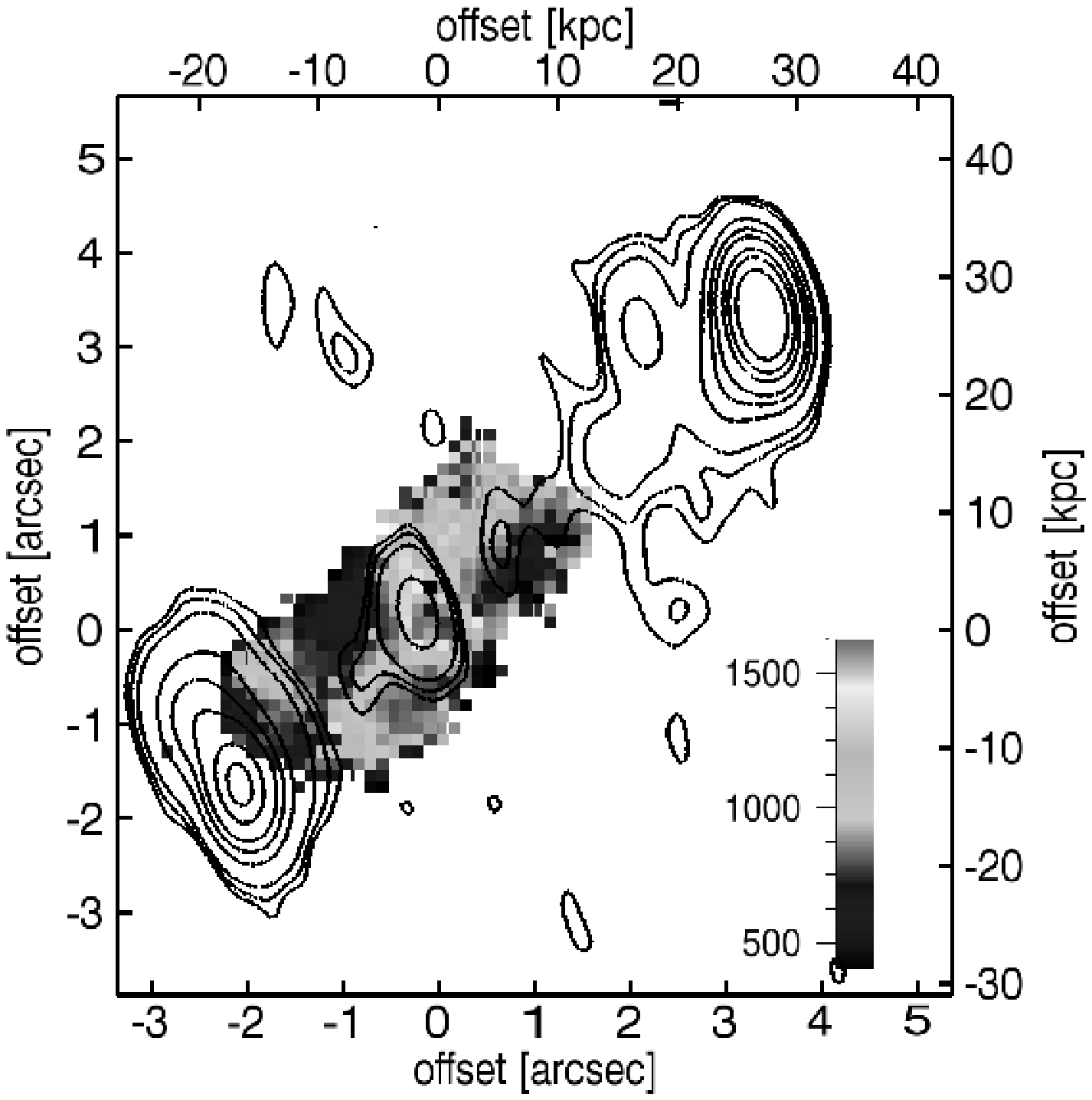,width=0.28\textwidth}
\caption{{\it left to right:} Emission-line morphology, and maps of
  relative velocities and line widths (Full widths at half maximum) of
  MRC0406-244 at z=2.4. For a color version of this figure see
  \cite{nesvadba08}.}
\end{figure}

Several arguments indicate that a 'cocoon' of hot, overpressurized gas
inflated by the radio jet may be the main driver of accelerating and
ionizing the dense and warm gas we observe in the optical line
emission. (1) Geometry: The major axes of the emission-line regions are
aligned with the axes of the radio jet and extend significantly
beyond the size of the stellar continuum. Although being elongated
along the jet axis, the morphology of the gas suggests a significant
lateral expansion, and not just a very localized interaction along the
jet working surface. The spatial extent of the emission-line regions
is smaller than the size of the radio lobes. Galaxies with compact
radio sources have compact line emission. (2) Timescales: From the
observed gas velocities and size of the emission-line regions we can
roughly estimate outflow times of few $\times 10^7$ yrs, similar to
typical AGN lifetimes, and significantly less than the typical
duration of a starburst. (3) Energy: Observed kinetic energies are
sufficient to unbind up to $\sim 10^{11}$ M$_{\odot}$ of gas from a
massive host galaxy \cite{nesvadba06}. Comparing the energy injection
rate of the AGN with the energy needed to drive the observed gas
kinematics, we find that about 10\% of the jet kinetic power are
needed to explain the gas motion. Using this efficiency and relating
the rest-frame energy equivalent of the supermassive black hole with
the kinetic energy of the outflows, we find that about 0.1\%$-$0.2\%
of the energy equivalent of the black hole is used to power the
outflows \cite{nesvadba08}. This is in good agreement with what is
assumed in models of galaxy evolution.

As discussed in \cite{nesvadba06,nesvadba08} it appears that these
outflows may terminate star formation in a way that is consistent with 
requirements from chemical evolution models. Subsequent accretion of
satellite galaxies observed around some z$\sim$2 radio galaxies is
likely not sufficient to significantly alter these properties
\cite{nesvadba08,hatch09}. 

Extended line emission with blueshifted and redshifted gas at high
redshift is often regarded as evidence for rotation, even in cases
where the low spatial resolution of the data make it very difficult to
robustly differentiate between abrupt velocity changes smeared out by
the seeing and intrinsically smooth velocity gradients indicative of
rotation
\cite{law07,wright08,starkenburg08,nesvadba08b,bournaud08,law09,forster09,epinat09}. Due
to the large spatial extend of the line emission in radio galaxies
this constraint is somewhat alleviated. Nonetheless it is interesting
to point out that the measured velocity offsets and sizes of the
emission-line regions would correspond to virial masses of up to
10$^{15}$ M$_{\odot}$ within a radius of $\sim$20 kpc, not correcting
for extinction. This is similar, or even exceeds, mass estimates for
the dark-matter halos underlying the overdensities of galaxies
observed around some radio galaxies (which may represent collapsing
protoclusters \cite{lefevre96,miley06,venemans07}) out to Mpc
scales. Thus the observed emission-line kinematics are clearly
supergravitational. Starbursts at high redshift do not produce much
higher emission-line surface brightnesses than strongly star-forming
regions at low redshift and are thus not sufficient to explain the
immense H$\alpha$ luminosities of our targets \cite{lehnert09} or the
observed mass loss rates in ionized gas \cite{nesvadba07b}. 

Comparison with cooling-flow clusters at low redshift 
does not suggest that the bright, turbulent emission-line regions we observe
in the rest-frame optical are related to massive cooling flows: In low
redshift clusters, the emission-line gas seems to avoid the radio
lobes, whereas at high redshift, emission-line regions and radio
emission are well aligned. Moreover, we find that the blueshifted
bubble is always on the side of the radio lobe with higher
polarization. If the Laing-Garrington effect approximately holds, this
suggests that the blueshifted gas is along the approaching radio lobe
indicative of a net outflow from the host galaxy \cite{nesvadba08}.

\begin{figure}
\epsfig{figure=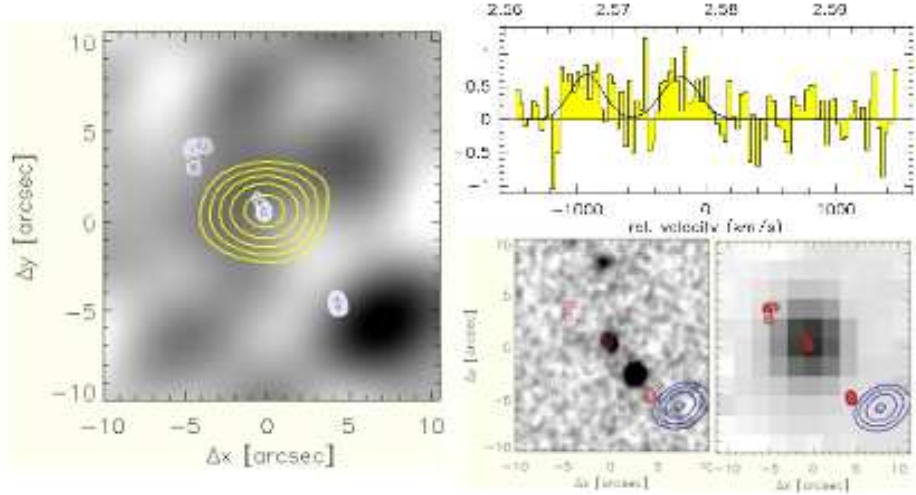,width=1.0\textwidth}
\caption{{\it left:} CO(3-2) line image of TXS0828+193 at z=2.6. We
  only detect the 3mm continuum at the position of the radio
  source (thick contours). Luminous CO(3-2) line emission originates from a region 90
  kpc SW from the radio galaxy, and near one of the radio hot
  spots (thin contour). The source is not spatially resolved at our 5 arcsec beam
  ($\sim$ 40 kpc). {\it top right:} The spectrum shows two components
  at redshifts which are in excellent agreement with that of the
  diffuse Ly$\alpha$ emission found at similar radii from the radio
  source. {\it bottom right:} We do not detect a counterpart at
  rest-frame UV to infrared wavelengths. The left panel shows an
  overlay with deep K-band imaging, the right panel shows our MIPS
  24$\mu$m image. Contours mark the position of the cm radio source and the CO line
  emitter, respectively. For color versions of these figures see \cite{nesvadba09}.}
\end{figure}

\section[AGN heating of the ICM in the early universe?]{Evidence for AGN heating of the intracluster medium in the early  
universe?}  
Our millimeter observations of CO line emission in z$\sim$2 radio
galaxies may also have yielded the first observational signature of
the deposition of parts of the mechanical energy of the radio jet into
the intracluster medium at high redshift \cite{nesvadba09}. To explain
the high entropy floors of X-ray luminous, massive galaxy clusters at
low and intermediate redshifts, models of galaxy evolution now
postulate a phase of efficient 'pre-heating' through the AGN at high
redshift, during the collapse of the cluster
\cite{nath02,mccarthy08}.

Using the IRAM Plateau de Bure Interferometer we detected faint
CO(3-2) line emission in the halo of the powerful radio galaxy
TXS0828+193 at z$\sim$2.6 which is at a distance of $\sim$90 kpc from
the radio galaxy, and does not seem associated with a significant
stellar mass greater than several $\times 10^9$ M$_{\odot}$
\cite{nesvadba09}. This is in stark contrast to all other galaxies at
similar redshifts detected in CO. We also do not detect the source in
24$\mu$m MIPS imaging, which covers the PAH bands in the rest-frame
mid-infrared, suggesting this gas is not forming stars at the
prodigious rates typically observed in massive, gas-rich high-redshift
galaxies.

This gas is near the hotspot of one of the radio jets of TXS0828+193, and we
detect two components with velocities that seem to 'straddle' that of the
faint, diffuse ionized gas observed in Ly$\alpha$ at similar radii from the
central radio galaxy \cite{villar02,villar03}. Given the atypical properties
of this CO line emitter, and proximity with the radio jet (spatially) and
diffuse gas (spectrally), we suspect we may have detected a cloud or filament
of cold gas, that is excited or may even be compressed by weak shocks from the
radio source.

Hydrodynamic jet models suggest that cloud collapse (and perhaps star
formation) may actually be enhanced through shocks driven by the
expanding radio source ('jet-induced star formation' or 'positive
feedback') \cite{mellema02,fragile04}. Extended filaments of cold
molecular gas have been observed in nearby X-ray luminous clusters
where they seem to be a by-product of interactions between the radio
jet and the cluster gas, which will ultimately heat the diffuse
cluster gas. More observations of similar stuctures will be necessary
to confirm this rough scenario of jet-induced gas collapse at high
redshift, which is based on deep observations with IRAM after a
recent, major upgrade, and a non-standard target selection. With the
advent of ALMA we will be able to further explore the many roles of
molecular gas in forming stars and shaping galaxies in the early
Universe.

\vspace{3mm}
{\it \noindent  NPHN acknowledges financial support through a fellowship of the Centre
Nationale d'Etudes Spatiales (CNES).}

\end{document}